\documentstyle[art12,fleqn]{article}
\def\soc{{\rm C}_{60}}
\def\rug{{\rm C}_{70}}

\def\la{\langle}
\def\ra{\rangle}
\def\beeq{\begin{equation}}
\def\eneq{\end{equation}}
\def\beeqa{\begin{eqnarray}}
\def\eneqa{\end{eqnarray}}

\addtocounter{section}{-1}
\begin{document}

\begin{center}

{\large {\bf{Long-Range Excitons in Optical Absorption Spectra\\
of Electroluminescent Polymer 
Poly({\mbox{\boldmath $para$}}-phenylenevinylene) } } }

\vspace{0.3cm}

{\rm Kikuo H{\sc arigaya}}\\
{\sl Physical Science Division, Electrotechnical Laboratory,\\
Umezono 1-1-4, Tsukuba 305, Japan}

\vspace{0.3cm}

(Received February 14, 1997)

\end{center}

\vspace{0.3cm}

\noindent
{\bf Abstract}

\vspace{0.3cm}

The component of photoexcited states with large spatial 
extent is investigated for poly({\sl para}-phenylenevinylene) 
using the intermediate exciton theory.  We find a peak due 
to long-range excitons at the higher-energy side of the lowest 
main feature of optical spectra.   The fact that the onset 
of long-range excitons is located near the energy gap is 
related to the mechanisms of large photocurrents measured in 
such energy regions.  We show that a large value of the
hopping integral is realistic for characterizing optical 
excitations.

\vspace{0.3cm}

\newpage

The remarkable electroluminescent properties of 
poly({\sl para}-phenylenevinylene) (PPV)$^{1)}$ have 
prompted physical and chemical research activities.  The 
polymer structure is shown in Fig. 1.  The onset of the 
photocurrent occurs at an excitation energy between 
3.0 eV and 4.0 eV,$^{\rm 2-4)}$ which is significantly 
larger than both of the optical absorption edge at about 
2.0 eV and the lowest peak energy at 2.4 eV.  The experiments 
have been interpreted theoretically in terms of excitonic effects 
which have been taken into account by the single excitation 
configuration interaction (single-CI) method$^{5,6)}$ and 
also by the density matrix renormalization group method.$^{7)}$  
The binding energy of excitons is about 0.9 eV, as  
estimated using the single-CI theory.$^{5)}$

The spatial extent of excitons, in other words, the 
distance between electrons and holes, depends on their 
photoexcitation energies.  If an exciton is strongly bound, 
its extent can be smaller than the region of the PPV monomer 
unit and thus the exciton becomes Frenkel-like, as observed 
in molecular crystals.  If the binding is weak, the 
photoexcited electron-hole pair tends to distribute over 
several monomer units, like the charge-transfer exciton in 
molecular systems.  The main purpose of this study is to 
characterize the extent of photoexcited states of the PPV 
using the single-CI theory recently developed by Shimoi and Abe.$^{6)}$
The polymer backbone structure is treated using a tight binding 
model Hamiltonian with electron-phonon interactions, and 
attractions between electrons and holes are taken into
account by long-range Coulomb interactions.  When the 
distance between an electron and a hole is shorter than 
the spatial extent of the monomer, we call the exciton
a ``short-range" exciton.  When the exciton width
is larger than the extent of the monomer, we call 
it a ``long-range" exciton.  We will characterize
each photoexcited state as ``short range" or ``long range"
by calculating the probability that the photoexcited 
electron and hole exist on different PPV monomer units.
A similar characterization method has also been used 
in recent investigations of charge-transfer excitons 
in $\soc$ cluster systems.$^{\rm 8-10)}$

We consider the following model with electron-phonon and 
electron-electron interactions.  
\beeqa
H &=& H_{\rm pol} + H_{\rm int}, \\
H_{\rm pol} &=& - \sum_{\la i,j \ra,\sigma} ( t - \alpha y_{i,j} )
( c_{i,\sigma}^\dagger c_{j,\sigma} + {\rm h.c.} ) 
+ \frac{K}{2} \sum_{\la i,j \ra} y_{i,j}^2, \\
H_{\rm int} &=& U \sum_{i} 
(c_{i,\uparrow}^\dagger c_{i,\uparrow} - \frac{n_{\rm el}}{2})
(c_{i,\downarrow}^\dagger c_{i,\downarrow} 
- \frac{n_{\rm el}}{2})  \nonumber \\
&+& \sum_{i,j} W(r_{i,j}) 
(\sum_\sigma c_{i,\sigma}^\dagger c_{i,\sigma} - n_{\rm el})
(\sum_\tau c_{j,\tau}^\dagger c_{j,\tau} - n_{\rm el}).
\eneqa
In eq. (1), the first term $H_{\rm pol}$ is the tight 
binding model along the PPV polymer backbone (shown in 
Fig. 1) with electron-phonon interactions which couple 
electrons with modulation modes of the bond lengths, and 
the second term $H_{\rm int}$ is the Coulomb interaction 
potentials among electrons.  In eq. (2), $t$ is the 
hopping integral between the nearest-neighbor carbon atoms 
in the ideal system without bond alternations; $\alpha$ 
is the electron-phonon coupling constant that modulates 
the hopping integral linearly with respect to the bond 
variable $y_{i,j}$ which is a measure of the magnitude of 
the alternation of the bond $\la i,j \ra$; $y_{i,j} > 0$ 
for longer bonds and $y_{i,j} < 0$ for shorter bonds (the 
average $y_{i,j}$ is taken to be zero); $K$ is the 
harmonic spring constant for $y_{i,j}$; and the sum is 
taken over the pairs of neighboring atoms.  Equation (3) describes
the Coulomb interaction among electrons.  Here, $n_{\rm el}$ 
is the average number of electrons per site; $r_{i,j}$ is 
the distance between the $i$th and $j$th sites; and 
$W(r) = 1/\sqrt{(1/U)^2 + (r/a V)^2}$
is the parametrized Ohno potential.  
The quantity $W(0) = U$ is the strength of the onsite 
interaction; $V$ means the strength of the long-range 
part ($W(r) \sim aV/r$ in the limit $r \gg a$); and 
$a = 1.4$\AA~ is the mean bond length.  We use the 
long-range interaction because the excited electron 
and hole are spread over a fairly large region of the system 
considered.  The parameter values used in this paper 
are $\alpha = 2.59t$/\AA, $K=26.6t$/\AA$^2$, 
$U=2.5t$, and $V=1.3t$ (ref. 6).  Most of the quantities 
in the energy units are shown by the unit of $t$ in this paper.

Excitation wavefunctions of the electron-hole pair are 
calculated by the Hartree-Fock (HF) approximation followed 
by the single-CI method.  This method has been used in the 
optical spectra of the PPV chain,$^{5,6)}$ and $\soc$ ($\rug$) 
molecules and solids.$^{\rm 8-11)}$  The optical absorption 
spectra become anisotropic with respect to the electric field 
of light, as expected from the polymer structures.  
Anisotropy effects are considered by applying an electric 
field in the direction parallel to the chain axis (shown
in Fig. 1) as well as in the perpendicular directions.

First, we show the total absorption spectra, and discuss the
effects of the boundary conditions and anisotropic effects 
with respect to the directions of the electric field of light.  
Figures 2 (a) and 2 (b) show the calculated spectra with 
periodic and open boundaries, respectively.  It seems 
that the number of PPV monomer units, $N=20$, gives 
the spectral shape which is almost independent of the chain 
length.  The optical spectra strongly depend on the system size 
when the number $N$ is less than 10, but they 
become almost independent of the size when $N$ is near 20.
Also, the difference in the spectral shapes of the two 
boundaries is small.  This is related to the 
saturated behavior of the spectral shape.

There are four features at 1.2$t$, 2.1$t$, 2.4$t$,
and 3.0$t$, in the total absorption spectra in Fig. 2.
Anisotropy effects with respect to the electric field
are clearly seen.  The first and third features are larger 
when the field is parallel to the polymer axis, while 
the second and fourth features are larger when the field is
perpendicular to the axis.  In the literature,$^{\rm 5-7)}$ 
there are two different schemes of theoretical 
assignment of absorption peaks.  Here, we have applied
both schemes  using $t=2.0$ eV (ref. 6) and $t=2.3$ eV 
(refs. 5 and 7).  The results are shown in Table I 
with the experimental absorption values of the 
chemically improved PPV reported in ref. 3.  The degree 
of agreement between experiment and theory is 
comparable for the two schemes in terms of
the optical excitation energies.  However, the recent 
measurement of anisotropy$^{12)}$ favors the assignment 
with $t=2.3$ eV.  The feature at around 4.7 eV shows 
reverse anisotropy from that of the lowest feature around 
2.5 eV.  The assignment of the experiment with $t=2.3$ eV
is consistent with the anisotropy shown in Fig. 2.  
Therefore, we find that the larger value of the hopping 
integral, $t=2.3$ eV, is realistic for optical excitations 
of the neutral PPV chain.

Next, we look at the origins of the main features in 
relation to the band structure.  Figure 3 shows the band 
structure obtained through the HF approximation.  The branches 
of the valence (conduction) band are labeled as VB$j$ (CB$j$) 
($j=1-4$) from the energy gap to the band edges.  The bands 
VB2 and CB2 almost lack dispersion.  This is due 
to the fact that the amplitudes at atoms A and D are 
nearly zero, and the wavefunctions are localized at  
atoms B, C, B', and C'.  Therefore, these two bands are 
nearly flat.  In the parallel electric field case, the 
dominant first and third features are mainly given by 
the transitions from VB1 to CB1 bands and from VB2 to CB2
bands, respectively.  In the perpendicular electric field 
case, the large second and fourth features originate from 
transitions from VB1 (VB2) to CB2 (CB1) bands and from 
VB2 (VB3) to CB3 (CB2) bands, respectively.

The long-range component is defined as below.  
First, we calculate the probability that the photoexcited 
electron and hole exist on different PPV monomer units.  
This probability is $1 - 1/N$ for the system with $N$ 
monomer units, when the electron and hole are distributed 
uniformly.  If it is less than $1 - 1/N$, the electron 
and hole tend to localize in a single monomer unit, and 
the excited state resembles a Frenkel exciton, which is 
typical for molecular crystals.  If the probability 
is higher than $1 - 1/N$, the electron and hole preferably 
separate into different monomer units, and the excited state 
has characteristics resembling those 
of a charge-transfer exciton in molecular 
crystals.  We call such an excited state the 
``long-range" component.  We can determine the contribution 
to the optical spectra from the long-range component by 
taking the sum over such states only.

Figure 4 shows the calculated optical absorption spectra
for the periodic boundary case.  The bold lines show the total
absorption spectra, and the thin lines the spectra from
the long-range component of excitons.  Figures 4 (a) and 4 (b)
show the absorption with the parallel and perpendicular
electric field cases, respectively.  In Fig. 4 (a), there is
a broad peak around the energy 1.7$t$ at the higher-energy
side of the feature at 1.2$t$.  At the higher-energy side
of the 2.4$t$ feature, the oscillator strengths of the 
long-range component are very small, and this is consistent 
with the origin of the 2.4$t$ feature wherein the VB2 and CB2 
bands in Fig. 3 are almost free of dispersions.  The optical 
excitations of this feature tend to localize on a single
PPV monomer.  In Fig. 4 (b), there are two peaks of the
long-range component around the energies, 2.3$t$ and 3.6$t$,
at the higher-energy side of the two main features
at 2.1$t$ and 3.0$t$.  Figure 4 (c) shows the orientationally
averaged spectra.  The spectral shape of the long-range
component is very broad and extends from 1.5$t$
to about 2.5$t$.  The threshold of the long-range component
is $1.566t$, which is slightly smaller than the energy of the 
HF energy gap 1.581$t$.  Such long-range excitons might 
play a part in the mechanisms of large photoconduction which 
has been measured in the energy region larger than the HF 
energy gap 1.581$t$.  The oscillator strengths of the 
long-range component are larger when the electric field is 
perpendicular to the chain axis.

Finally, we calculate the ratio of the sum of oscillator strengths
of the long-range excitons with respect to that of the total 
absorption, in other words, the ratio of the area below the 
thin line with respect to the area below the bold line in 
Fig. 4.  The quantity is calculated for each case of the 
electric field direction, and the results are shown as a 
function of $N$ in Fig. 5.  The ratio strongly depends on 
the number of monomer units when $N$ is smaller than 10, due to
the finite system size.  However, the size dependence becomes
smaller as $N$ increases.  The ratio seems to be almost 
constant at about 0.08 near $N=20$.  The ratio
is anisotropic with respect to the direction of the electric
field of light.  The numerical data for the parallel and
perpendicular electric field cases are also shown in Fig. 5.
The ratio for the perpendicular field case
is slightly larger than that for the parallel field case.
This is related to the fact that the 2.4$t$ feature 
in Fig. 4 (a) has a very small long-range component
and the electron and hole around this feature tend to 
localize on a single PPV monomer.

In summary, we have investigated the component of photoexcited 
states with large spatial extent in the PPV chain.  The energy 
position of long-range excitons is nearly the same as the 
HF energy gap.  This finding is related to the mechanisms 
of large photoconduction measured in such energy regions.  Next, 
we calculated the ratio of oscillator strengths due to 
long-range excitons with respect to the sum of all the oscillator 
strengths of the absorption as a function of the PPV monomer 
number.  The ratio strongly depends on the system size when the 
monomer number is small, but becomes almost constant when the 
monomer number is more than 10 and near 20.

\mbox{}

\begin{flushleft}
{\bf Acknowledgements}
\end{flushleft}

Useful discussions with Y. Shimoi, S. Abe, S. Kobayashi, 
K. Murata, and S. Kuroda are acknowledged.  Helpful 
communication with S. Mazumdar, M. Chandross, W. Barford, 
D. D. C. Bradley, R. H. Friend, E. M. Conwell, and 
Z. V. Vardeny is greatly appreciated.

\pagebreak

\noindent
{\bf References}

\vspace{0.3cm}

\noindent
1) J. H. Burroughes, D. D. C. Bradley, A. R. Brown, R. N. Marks,
K. Mackay, R. H. Friend, P. L. Burn and A. B. Holmes: Nature 
{\bf 347} (1990) 539.\\
2) K. Pichler, D. A. Haliday, D. D. C. Bradley, P. L. Burn,
R. H. Friend and A. B. Holmes: J. Phys.: Condens. Matter 
{\bf 5} (1993) 7155.\\
3) D. A. Halliday, P. L. Burn, R. H. Friend, D. D. C. Bradley,
A. B. Holmes and A. Kraft: Synth. Met. {\bf 55-57} (1993) 954.\\
4) K. Murata, S. Kuroda, Y. Shimoi, S. Abe, T. Noguchi
and T. Ohnishi: J. Phys. Soc. Jpn. {\bf 65} (1996) 3743.\\
5) M. Chandross, S. Mazumdar, S. Jeglinski, X. Wei,
Z. V. Vardeny, E. W. Kwock and T. M. Miller: 
Phys. Rev. B {\bf 50} (1994) 14702.\\
6) Y. Shimoi and S. Abe: Synth. Met. {\bf 78} (1996) 219.\\
7) W. Barford and R. J. Bursill: in {\sl Proceedings of the 
International Conference on Science and Technology of Synthetic 
Metals (ICSM96)} (Elsevier, Netherlands, 1996).\\
8) K. Harigaya and S. Abe: Mol. Cryst. Liq. Cryst. {\bf 256} 
(1994) 825.\\
9) K. Harigaya and S. Abe: in {\sl 22nd International Conference
on the Physics of Semiconductors} (World Scientific, Singapore,
1995) p. 2101.\\
10) K. Harigaya: Phys. Rev. B {\bf 54} (1996) 12087.\\
11) K. Harigaya and S. Abe: Phys. Rev. B {\bf 49} (1994) 16746.\\
12) M. Chandross: Doctor Thesis (University of Arizona, 1996).\\

\pagebreak

\noindent
Table I.  The excitation energies of the main four features.
The energy value is not the onset of each feature,
but the peak energy.  The experimental values are
taken from the optical spectra of the 
chemically improved PPV in ref. 3.

\mbox{}

\begin{tabular}{ccc} \hline
experiment (ref. 3) & theory ($t=2.0$ eV) & theory ($t=2.3$ eV) \\ \hline
2.5 (eV)            & 2.4 (eV)            & 2.8 (eV) \\
3.8                 & 4.2                 & --      \\
4.7                 & 4.8                 & 4.8     \\
5.8                 & 6.0                 & 5.5     \\
--                  & --                  & 6.9     \\ \hline
\end{tabular}

\pagebreak

\begin{flushleft}
{\bf Figure Captions}
\end{flushleft}

\mbox{}

\noindent
Fig. 1.  The structure of the PPV chain.  The direction 
of the polymer chain axis is indicated by the arrow.  
The eight carbon sites, not equivalent to each other, are 
labeled A-F, B', and C'.

\mbox{}

\noindent
Fig. 2.  Optical absorption spectra of the PPV for
(a) periodic and (b) open boundaries.   The number of 
PPV units $N$ is 20.  The bold line indicates the total
absorption.   The thin (dotted) line indicates the 
absorption where the electric field is along  
(perpendicular to) the polymer axis.  The Lorentzian 
broadening of $\gamma = 0.15 t$ is used.

\mbox{}

\noindent
Fig. 3.  The band structure of the PPV in the Hartree-Fock
approximation.  There are four branches (VB$j$, $j=1-4$)
in the valence band, and four (CB$j$, $j=1-4$) in the 
conduction band.  Only the wavenumber region $0 < k < \pi$ 
is shown because of symmetry.  The lattice constant of 
the unit cell is taken as unity.

\mbox{}

\noindent
Fig. 4. Optical absorption spectra of the PPV with periodic 
boundary.  The polymer axis is in the $x$-$y$ plane.   The 
electric field of light is parallel to the chain and
in the direction of the $x$-axis in (a), and it is
perpendicular to the axis and along the $z$-axis
in (b).  The orientationally averaged spectra are shown
in (c).   The number of PPV units $N$ is 20.  The bold 
line indicates the total absorption.   The thin line indicates 
the absorption of the long-range component.  The Lorentzian 
broadening of $\gamma = 0.15 t$ is used.

\mbox{}

\noindent
Fig. 5.  Long-range component of the optical absorption
spectra as a function of the PPV unit number $N$ for the 
case with periodic boundary.   The squares are for the 
total absorption.  The circles and triangles indicate the
data for the cases with the electric field parallel and
perpendicular to the polymer axis, respectively.

\end{document}